\documentclass[letter]{aa}
\usepackage{graphicx}
\usepackage{bm}
\usepackage{natbib}
\usepackage{txfonts}

\begin{document}
\bibliographystyle{aa}
\input epsf

\title{Solar atmospheric oscillations and the chromospheric magnetic topology}

\author{A. Vecchio\inst{1}
\and G. Cauzzi\inst{1}
\and K. P. Reardon\inst{1}
\and K. Janssen\inst{1}
\and T. Rimmele\inst{2}}

\institute{INAF - Osservatorio Astrofisico di Arcetri, 50125 Firenze, Italy
\and National Solar Observatory, P.O. Box 62, Sunspot NM 88349, USA}
\date{\today}

\abstract{} 
{We investigate the oscillatory properties of the quiet solar
chromosphere in relation to the underlying photosphere, with
particular regard to the effects of the magnetic topology.} 
{For the first time we perform  a Fourier analysis on a sequence of
line-of-sight velocities measured simultaneously in a photospheric (Fe I
709.0 nm) and a chromospheric line (Ca II 854.2 nm). The velocities were
obtained from full spectroscopic data acquired at high spatial resolution
with the Interferometric BIdimensional Spectrometer (IBIS). The  field of view 
encompasses a full
supergranular cell, allowing us to discriminate between areas with different
magnetic characteristics.} 
{We show that waves with frequencies above the
acoustic cut-off propagate from the photosphere to upper layers
only in restricted areas of the quiet Sun. A large fraction of the
quiet chromosphere is in fact occupied by ``magnetic shadows'', 
surrounding network regions, that we identify as originating from
fibril-like structures observed in the
core intensity of the Ca II line. 
We show that a large fraction of the chromospheric acoustic power at 
frequencies below the acoustic cut-off,
residing in the proximity of the magnetic network elements, directly 
propagates from the
underlying photosphere. This supports recent results arguing that network
magnetic elements can
channel low-frequency photospheric oscillations into the chromosphere, thus
providing a way to input mechanical energy in the upper layers.}

\maketitle
\titlerunning{}
\authorrunning{Vecchio et al.}

\section{Introduction}

The properties of acoustic waves propagating in a magnetized 
solar atmosphere have received considerable attention. Recent efforts include the
utilization of UV diagnostics, in the form of time-series of both 1-D
spectra (SUMER) and 2-D continuum intensities (TRACE), to address
the behavior of  chromospheric oscillations \citep[see,
e.g.][and references therein]{judge_06}. Such studies have highlighted the
presence of areas called ``magnetic shadows'' surrounding quiet-Sun 
magnetic network elements and 
lacking both brightness and oscillatory power in the $2-3$ minute range
\citep{judge_01,krijger_01}. They have been attributed
to the interaction of acoustic oscillations with the local  magnetic field,
in particular to the effects on wave mode propagation or conversion
introduced by a highly structured magnetic configuration at  chromospheric
levels \citep{mcintosh_01}. Using magnetic and velocity field measurements
from MDI, it has further been determined  that oscillations are suppressed
when the diagnostics employed originate above the layer where the
plasma $\beta$ (the ratio of plasma pressure to magnetic pressure)
equals unity \citep{mcintosh_03}. This is the layer of the
``magnetic canopy'' that separates the weakly magnetized photosphere from
higher regions, in which magnetic field dominates the dynamics of the
plasma. Study of the shadows might help in evaluating the
relative contributions of magnetic heating in different areas of the quiet
Sun, and in assessing the role of high frequency (above the acoustic
cutoff of about 5 mHz) versus lower frequency oscillations \citep{judge_01}. 
 
Much discussion has also been devoted to  the `leakage' of
photospheric oscillations and flows into the chromosphere, due to
a lowering of the acoustic cut-off frequency within
inclined magnetic fields \citep{depontieu_04,hansteen_06,mcintosh_06}.
While these analyses have mostly been based on observations of
plage or active regions, it can be expected that the same
kind of effect would be observable in the network fields of the
quiet Sun, provided one can reach the necessary
spatial and temporal resolution. Indeed,  
recent results reported in \citet{jefferies_06} clearly
point in this direction, showing that a sizable fraction of the
photospheric acoustic power at frequencies below the cut-off 
might propagate to higher layers within and around
the quiet magnetic network elements. The propagating waves could represent a
significant source of energy to heat 
the ambient solar chromosphere \citep{judge_04,mcintosh_06}.

In this paper we analyze a time series of chromospheric and
photospheric velocities, obtained at high spatial resolution over an
extended field of view (FOV). The data were acquired
with the Interferometric Bidimensional Spectrometer
\citep[IBIS,][]{cavallini_06,cavallini_reardon_06}, installed at
the Dunn Solar Telescope of the {US} National Solar Observatory.
IBIS combines the
advantages of a full spectroscopic analysis on atmospheric periodicities,
usually performed with single-slit spectrographs, with the high spatial
resolution, high temporal cadence and FOV typical of filter
instruments. Observations of lines originating in
widely spaced layers of the solar
atmosphere allow a fairly direct estimation of the propagating character 
of waves.

\section{Observations and data analysis} \label{s_obs}

IBIS is used together with high-order adaptive optics
\citep{rimmele_04}, and provides quasi-monochromatic images 
in the range 580-860 nm (FWHM = 2 -- 4.5 pm). The spectral passband 
is sequentially stepped through
multiple wavelengths in a number of selected lines, providing full spectral
information over a circular FOV 80$''$ in diameter. 

\begin{figure}
\hbox{
\hglue -1.0cm
\includegraphics[width=5.8cm,height=5.8cm]{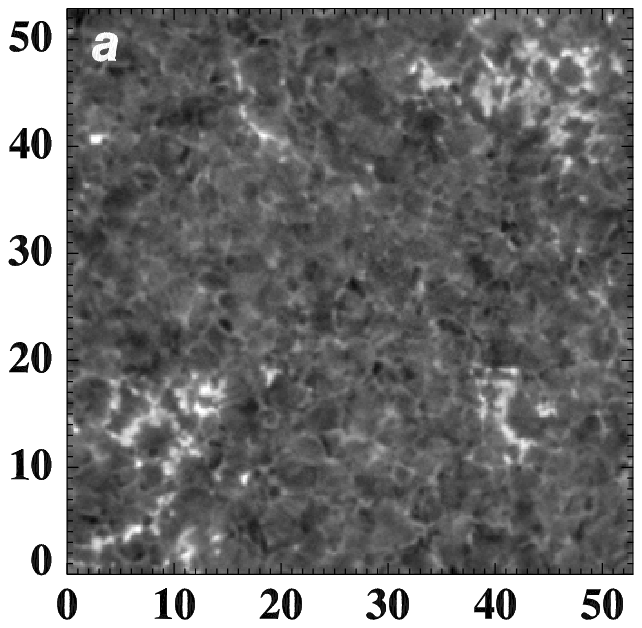}
\hglue -2.05cm
\includegraphics[width=5.8cm,height=5.8cm]{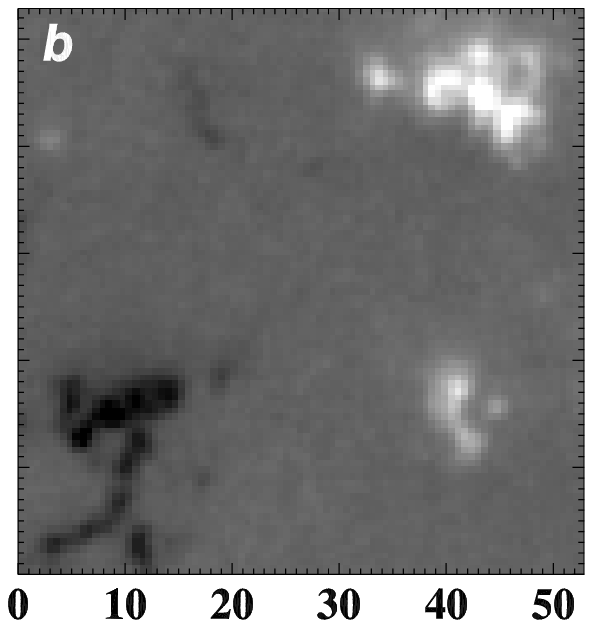}}
\caption{Region under analysis in the middle of the sequence.
FOV is $53'' \times 53''$ in area (0.17$"$ pixel
scale). Tick 
marks in arcsec.
{\it a)} Intensity image at $\Delta\lambda=0.16$ nm from Ca II 854.2 line center.
The intensity at this wavelength is very sensitive to 
the presence of magnetic structures \citep[cf.][]{leenaarts_06}.
{\it b)} co-temporal HR MDI map showing a mixed polarity, enhanced network area. 
The map is saturated at $\pm$ 200 G.
The higher resolution IBIS image clearly
reveals how single magnetic features visible in MDI are actually composed
of several distinct structures of sub-arcsec size.}
\label{fig_fov}
\end{figure}

\begin{figure*}
\vbox{
\hbox{
\hglue -0.6cm
\includegraphics[width=6cm,height=6cm]{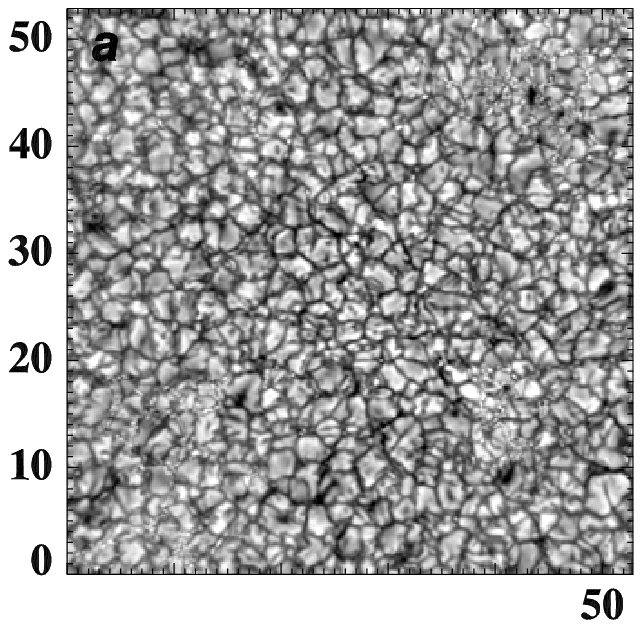}
\hglue -2.15cm
\includegraphics[width=6cm,height=6cm]{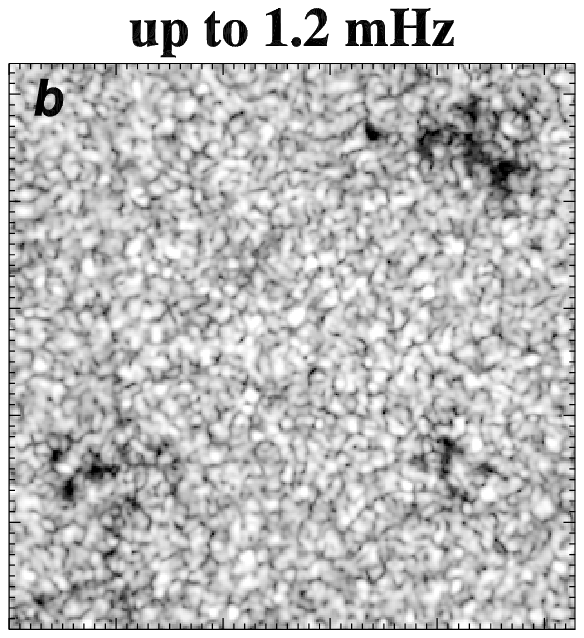}
\hglue -2.15 cm
\includegraphics[width=6cm,height=6cm]{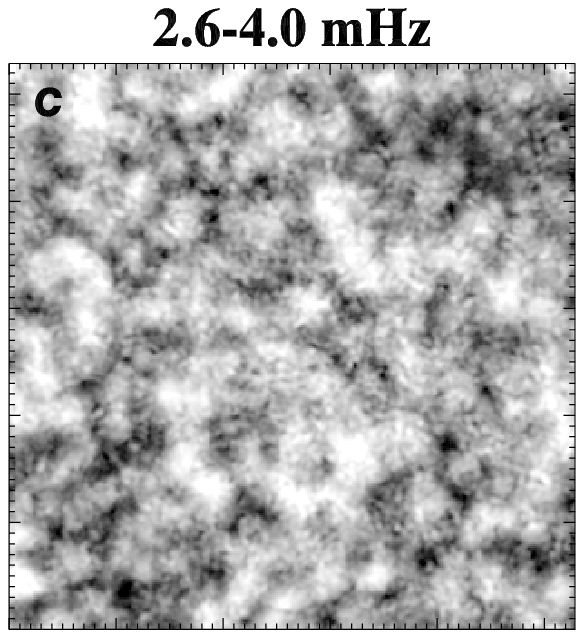}
\hglue -2.15 cm
\includegraphics[width=6cm,height=6cm]{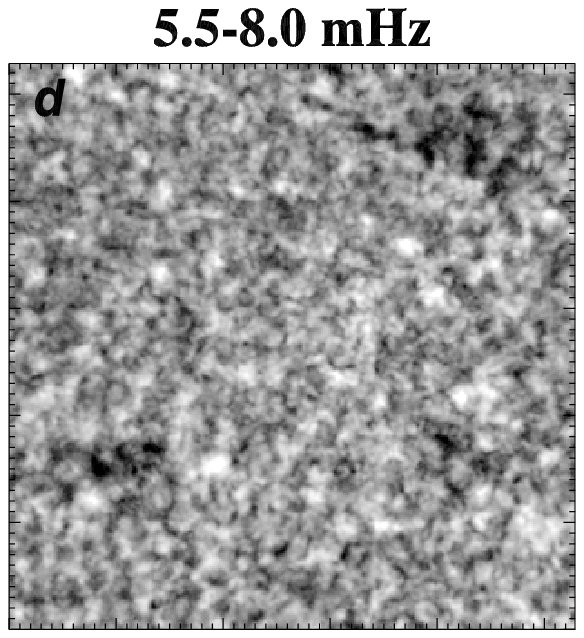}
}
\vglue -2.1cm
\hbox{
\hglue -0.6cm
\includegraphics[width=6cm,height=6cm]{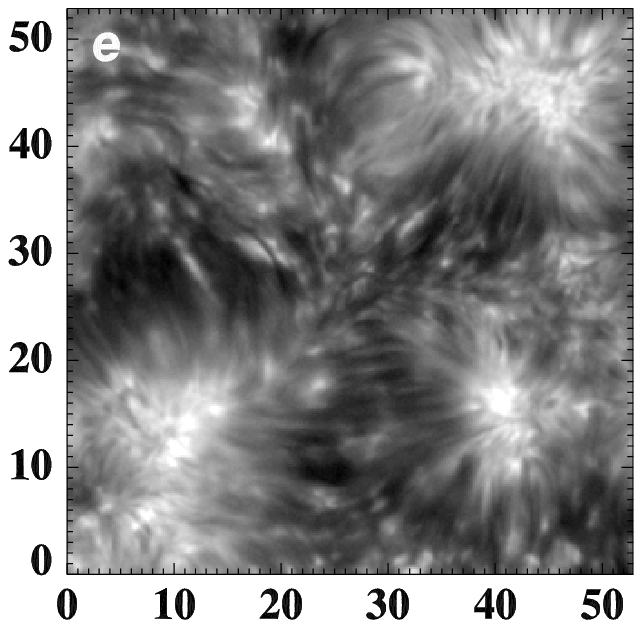}
\hglue -2.15cm
\includegraphics[width=6cm,height=6cm]{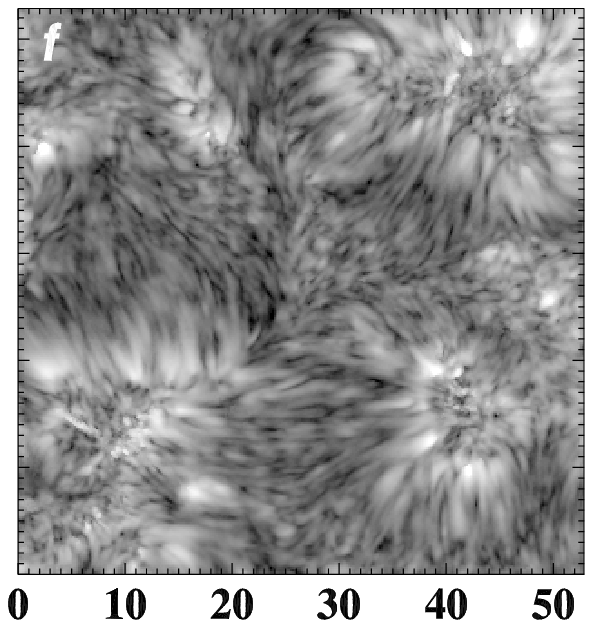}
\hglue -2.15cm
\includegraphics[width=6cm,height=6cm]{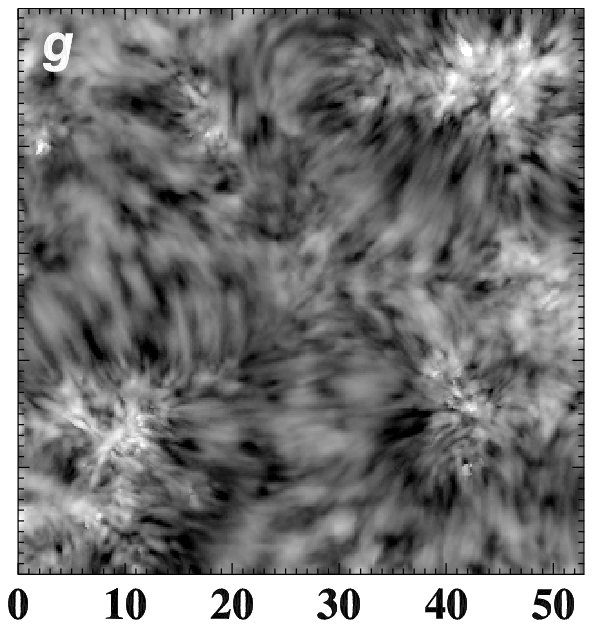}
\hglue -2.15cm
\includegraphics[width=6cm,height=6cm]{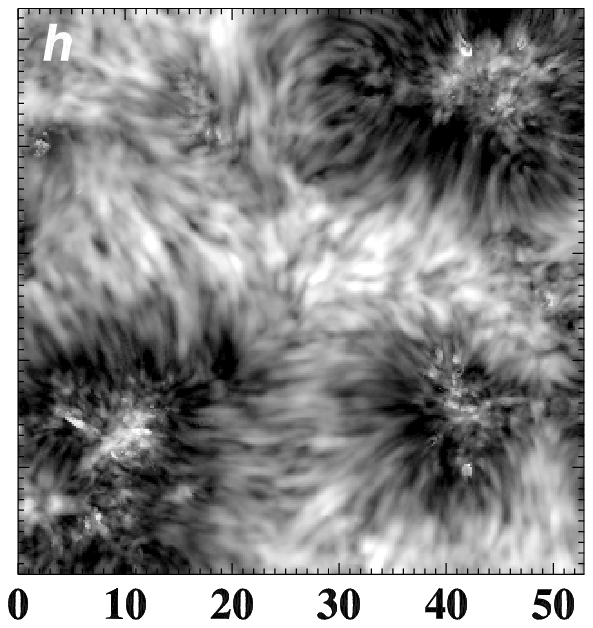}
}
}
\caption{
Panel {\it a}: Speckle reconstructed broadband continuum at 710 nm for the
FOV under analysis. The area 
outlines a supergranular
cell with quiet granulation in the center and some abnormal granulation at
the edges. 
Panel {\it e}: Line core intensity of CaII 854.2 nm
showing {fibrils}
outlining the chromospheric canopy. 
Panels {\it b--d} and {\it f--h}: Spatially resolved Fourier's velocity power 
maps, calculated over the full observing period, and
averaged over the range of frequencies indicated.  The intensity
scale is logarithmic, with bright indicating larger Fourier amplitudes.
Each image is normalized for optimal display. 
Panels {\it b--d}: photospheric Fe I 709.0 nm line. In the non-magnetic
areas, the low frequency map displays enhanced power at the
spatial scales typical of granulation, while in the intermediate
frequencies larger scales  of $6-8"$ associated with the
$p$-modes are recovered. Panels {\it f--h}: chromospheric Ca II 854.2 nm 
line.}
\label{fig_pow}
\end{figure*}

The data utilized here were acquired on June 02, 2004, in a quiet region at
disk center \citep[]{janssen_06}. We focus on data acquired in the
mid-photospheric Fe I 709.0 nm line and in the low-chromospheric Ca II
854.2 nm line, that were sampled respectively every 4 s and 7 s (16 and 27
spectral positions per line). An {approximately one-hour-long series} was obtained
under good seeing conditions, with a repetition rate of 19 s and a pixel
scale of 0.165$''$ per pixel. The data were corrected for instrumental
effects, tracked in time to remove any drifts in the AO lock point (set on
the granular field), and aligned to a common FOV between the two
wavelengths.  The final analysis was performed on the central portion of
the original FOV,  shown  at different wavelengths  in Figs.~\ref{fig_fov}
and \ref{fig_pow} (left panels). The
spatially and temporally averaged spectral profiles well reproduce the
atlas disk-center intensity profiles for both lines. 
Of particular interest are the {fibrils} visible in the Ca II
line core intensity (Fig. \ref{fig_pow}{\it e}) {which fan out around the
network into the surrounding internetwork regions.
These fibrils exhibit only a gradual evolution with no evidence of rapid 
variation at our resolution.}

Standard Fourier analysis, including retrieval of phase difference and coherence
spectra, was then applied to the temporal series of line-of-sight
velocities. The latter were calculated at full spatial resolution in both lines as the
position of  the minima of the spectral profiles. Other analysis tecniques 
might be more applicable to the study of the spatio-temporal dynamics, but 
Fourier analysis provides an initial insight into the chromospheric behavior 
and indicates areas for further investigation. The spatially averaged
velocity power spectra for the two lines are consistent with earlier results
obtained using single-slit spectrographic observations of quiet solar
regions \citep[e.g.][]{fleck_89}: the photospheric line shows a sharp peak
around $3.5$ mHz due to the (evanescent) 5-min oscillations, while the
chromospheric line has a broader distribution of power  with a significant
amount of power distributed {near the $5.5$ mHz} cut-off.

The spatial distribution of the velocity power is displayed in Fig.
\ref{fig_pow} (panels {\it b}--{\it d} and {\it f}--{\it h}),  averaging
the power spectra over three selected temporal frequency bands: up to 1.2
mHz (the ``evolutionary'' range); 2.4 -- 4.0 mHz (the ``evanescent'' range);
and 5.5 -- 8.0 mHz (the ``high frequency'' range). Unlike earlier analyses
\citep[]{mcintosh_01,mcintosh_03}, we have found it illuminating for our
high resolution data to differentiate between the last two frequency
ranges,
as will be shown below.
Fig. \ref{fig_phases}
displays the spatial distribution of the phase difference {spectra} between the
photospheric and chromospheric velocities, {calculated over $5\times 5$
spatial pixels, and} averaged over the evanescent 
and high frequency ranges. Phase difference signals at low frequencies
have a salt-and-pepper appearance throughout the FOV without outlining any
obvious structure, and are not analyzed in this paper.

\section{Results}\label{s_results}

The photospheric velocity power maps (Fig. \ref{fig_pow}{\it
b}, {\it c}, {\it d}) display well known characteristics such as lack
of power in the network elements at all frequencies, especially for the
locations with stronger and more spatially coherent magnetic elements 
\citep[]{thomas_00,muglach_05}. This suppression is clearly observable, 
especially for the higher
frequencies, only when the spatial resolution is of the order of 1$''$ 
or higher.
In the following we focus on the Ca II power maps 
(Fig. \ref{fig_pow}{\it f}, 
{\it g}, {\it h}), as they provide a
strikingly different picture: at all frequencies one can discern 
filamentary structures, likely outlining the magnetic canopy, 
that fan out from the network elements, 
and reach
towards the quiet internetwork or (apparently)
connect regions of opposite
polarity. Their shape, position, and general appearance readily recall
the loop structures visible in the core of the CaII 854.2 nm (cf. Fig.
\ref{fig_pow}{\it e}), whose presence thus defines different areas 
of the FOV with very diverse oscillatory character.  

{\it Evolutionary range:} At low frequencies (Fig. \ref{fig_pow}{\it f})
about  70\%
of the  total velocity power is contained within the inner boundary 
of fibrilar structures that accounts
for  only 25\% of the pixels. The highest power patches are located nearby,
but not coinciding with, the actual photospheric
magnetic network points  
-- compare, e.g., the very strong feature at position (38,18). We note however
that in data of limited spatial resolution such a distinction might be
difficult to appreciate, and this could be the cause of earlier reports of
long period (longer than $\approx$ 10 minutes) oscillations 
within the chromospheric network
\citep[e.g.][]{mcateer_02}. 

\begin{figure}[h]
\hbox{
\hglue -1.0cm
\includegraphics[width=5.8cm,height=5.8cm]{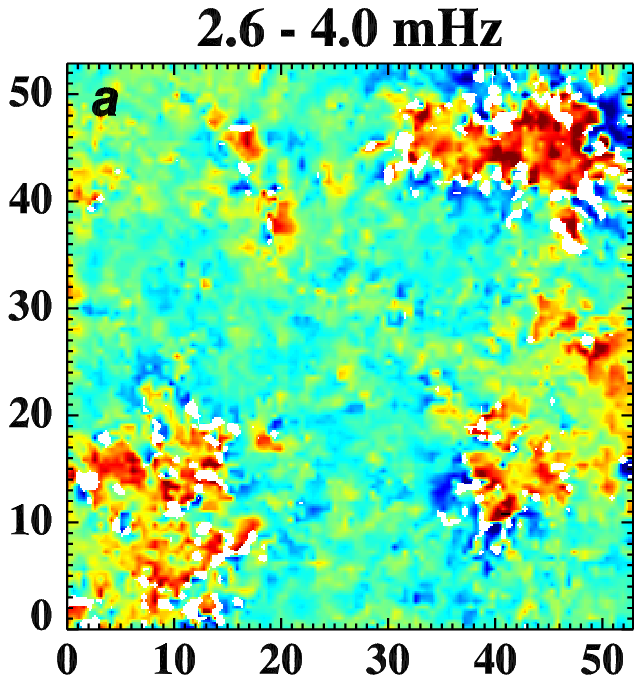}
\hglue -2.05cm
\includegraphics[width=5.8cm,height=5.8cm]{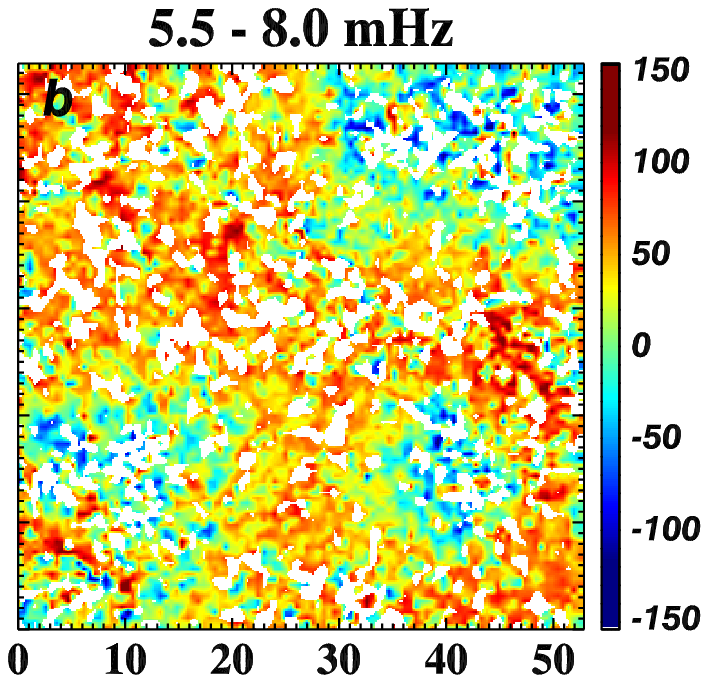}}
\caption{Phase difference maps between the photospheric and
chromospheric velocities for the evanescent {({\it a})} and high 
frequency {({\it b})} ranges of
Fig. \ref{fig_pow}. {White regions have coherences below the 
limit of $0.7$ ({\it a})/ $0.6$ ({\it b}) calculated from a 
random spatio-temporal field.}
Positive, red values indicate upward travelling
disturbances. Color scale in degrees, tickmarks in arcseconds.
}
\label{fig_phases}
\end{figure}

{\it Evanescent regime:} At intermediate frequencies (Fig. 
\ref{fig_pow}{\it g})  
significant chromospheric power is found within the ``fibril-free'' 
internetwork 
region. Fig. \ref{fig_phases}{\it a} shows that
phase differences in this region have small values, thus confirming 
the essentially evanescent
character of the waves. In this area the coherence is spatially uniform 
with values greater than 0.9. However, the most interesting feature is the
strong concentration of power occurring within and immediately around the
magnetic network (Fig. 
\ref{fig_pow}{\it g}), with about 5\% of the pixels making up 20\% of the
signal. In these areas the phase difference has large positive values, 
reaching
coherences up to 0.9. These results strongly support the findings 
of \citet{jefferies_06}, who argue that
leakage of low-frequency photospheric oscillations into the chromosphere through
`magneto-acoustic portals', positioned within the network,
might provide a significant source of the energy
necessary for heating the quiet chromosphere.

The higher spatial resolution of our data, coupled with a larger
separation between lower and upper layers, allows us further insights into
this phenomenon \citep[the two spectral lines  used by ][ have 
an average separation of only 250 km]{jefferies_06}.  For example,
Fig. \ref{fig_pow}{\it g} shows
that in several instances the highest chromospheric power is
located at the {\it edge} of  magnetic elements  rather than directly
within them - compare the elongated structure around position
(9,13). This is probably due to a significant inclination of the
magnetic fields in some part of the network, so that the
chromospheric signal originating in the magnetic footpoints will appear
displaced with respect to the source. We note that the same effect will
also cause a loss of apparent spatial connection 
between the photospheric and
chromospheric velocities, perhaps justifying the frayed appearance of both
phase and coherence maps in these areas, as well as the lower coherence
values. We will examine {this issue} in more detail in a future work.

{\it High frequency range.} The power map of
Fig. \ref{fig_pow}{\it h}, obtained for frequencies above the acoustic cut-off,
 provides a clear insight into the `magnetic shadows' described in the
Introduction. 
{The region affected by the presence of the Ca II fibrils}
shows a much reduced
oscillatory power (about 25\% on average)
with respect to the remaining `quiet' FOV. 
From comparison with the previous figures it is clear that the distinction in
oscillatory properties is
based not on the photospheric magnetic field but on its
3-D topology: most of the area lying below the canopy displays normal
granulation, and lacks any particular signature in MDI magnetic maps. 

The phase difference map of Fig. \ref{fig_phases}{\it b} also shows 
a `segregated'
behavior, with consistently positive values only in the regions of
hig-power.  We take this as indication that acoustic disturbances
can propagate to chromospheric
layers only when not impeded by the presence of the canopy that can absorb,
scatter or transform them by means of a strongly inclined magnetic field
\citep{bogdan_03, schunker_06}. 

The {spatially averaged} coherence is approximately 0.6 in the 
regions of positive phase
difference, and less than 0.4 in the {``fibrilar"} areas 
{around the network} \citep[a fact
already noted in][]{deubner_90}.
 However, while the spatial distribution of
the coherence in the canopy {region} is rather compact, it is very 
fragmented in
the central part of the FOV, with patches of coherence as high as 0.9 
lie next to ones with very low values (white patches 
in Fig. \ref{fig_phases}{\it b}) We believe this is due, at least in
part, to the presence  of acoustic shocks 
 that start developing at the typical height of
formation of the Ca II line \citep[]{carlsson_97}.

\section{Discussion and conclusions}\label{s_discussion}

Our high resolution velocity power and phase difference maps strongly support the
view of \citet[]{mcintosh_01} that even the quiet solar atmosphere can be
sharply partitioned in regions of different `connectivity' between the
lower photosphere and higher layers, depending on the local magnetic
topology. Our findings are consistent with
earlier results obtained with the analysis of chromospheric UV continua
\citep{judge_01,mcintosh_03}. Despite the added
complexity of line formation in a highly structured environment, it appears
that the
Doppler shift of the Ca II 854.2 nm line can be employed as a reliable
diagnostic of the low-chromosphere dynamics. Even with the restrictions due
to terrestrial seeing and sequential spectral acquisition, our data
provide both higher spatial resolution and a clearer diagnostic of
chromospheric dynamics than other observations, for example by
TRACE. In particular,
the latter is  due to the fact that our signatures form in layers  much
more widely separated than the UV continua at 160 and 170 nm commonly used.

Our results suggest that the highly dynamic fibril-like structures visible 
in the core of
the CaII 854.2 nm line can safely be taken as proxies of the magnetic
canopy that threads the chromosphere. These structures seem to represent an
atmospheric component that until now has been mostly overlooked, but that 
has the potential to fill many gaps in our understanding of the
chromosphere or even reconcile (apparently) conflicting results. For example, 
our data readily explain why comparable quiet Sun chromospheric
observations taken with slit spectrographs  sometimes
provide confusing views \citep{judge_01,pietarila_06}.
Most often, a spectrograph's slit is positioned on the basis
of photospheric signatures, such as MDI longitudinal field maps, brightness
in the CaII wings, etc. in order to avoid magnetic strutures. However, as
clearly shown in Figs. \ref{fig_pow}{\it h} and \ref{fig_phases}{\it b}, 
the observed chromospheric
properties strongly depend on the presence of {fibrilar regions}, whose 
location and
extension is not immediately predictable from photospheric maps.

Fig. \ref{fig_pow}{\it h} shows that an appreciable amount of
chromospheric power in the 3-minute range is restricted to a limited fraction of
the field of view. This area is much smaller than what would be defined  as
``internetwork'' on the basis of photospheric signatures -- for the data
discussed  here this fraction is about 50\%. 
Hence the issue of what
actually heats the chromosphere becomes even more pressing, as the
(non-magnetic) acoustic wave dissipation theory might apply to a less `quiet
Sun' than conventionally assumed. A possible solution is
provided by the power and phase difference maps of Fig. \ref{fig_pow}{\it
g} and Fig. \ref{fig_phases}{\it a}
that show how the network magnetic elements can channel photospheric acoustic 
power into upper layers 
at frequencies below the cut-off. As noted in
\citet[]{jefferies_06}, the  net mechanical energy flux provided by these waves could provide a considerable
fraction of the energy needed to balance the radiative chromospheric
losses. Given the complexity in the actual atmospheric stratification, we will not
calculate this energy.
We note instead that our spectrally resolved observation would be 
suited for a detailed comparison
with hydrodynamical simulations of the kind presented in
\citet{hansteen_06}. In particular, at the heights sampled by
the 854.2 nm line such
propagating waves might be already developing into the slow shocks invoked
by \citet{depontieu_04}, thus providing a way to deposit the energy in the
quiet chromosphere. 

It is possible that  the small fraction of  the FOV possessing
significant 3-minute power is a result of the particular area analyzed in
this paper, in which the enhanced, bipolar network elements  might alter
the local field topology in ways not completely representative of the quiet
Sun.  While acknowledging that there might be differences with
respect to supergranular cells enclosed by weak unipolar network 
(or, even more so, cells in coronal holes), the active network is normally 
present across the
Sun and the cycle, as a review of MDI data show. Our temporally and spatially 
averaged data are fully consistent with
both atlas intensity profiles and previously published averaged power
spectra. Nevertheless, a survey of oscillatory behavior in various quiet-Sun
configurations would be very revealing, especially if coupled with
simultaneous TRACE observations to confirm the shadow extensions. We plan to
perform coordinated observations of this kind in the near future.

\acknowledgements{
We are grateful to P. Cally, B. Fleck, and F. Hill for
comments and discussions. F. W\"oger kindly provided the speckle
reconstruction code. We thank the referee Rob Rutten for the detailed
comments. IBIS was built with contributions from
INAF/Arcetri
Observatory, the University of Florence, the University of Rome Tor
Vergata, and MIUR. NSO is operated by the
Association of Universities for Research in Astronomy,  Inc. (AURA), under
cooperative agreement with the National Science Foundation.  This research
was partially funded through the European Solar Magnetism Network (ESMN,
contract HPRN-CT- 2002-00313), PRIN-MIUR 2004 and the Italian Ministry
of Foreign Affairs (MAE)}

\bibliography{shadows}

\end{document}